\documentclass{article}       
\usepackage{e}                
\usepackage{graphicx}
\def \bea{\begin{eqnarray}}
\def \beq{\begin{equation}}
\def \b{{\cal B}}
\def \bb{\overline{{\cal B}}}
\def \eea{\end{eqnarray}}
\def \eeq{\end{equation}}
\def \gb{\overline{\Gamma}}
\def \gs{\stackrel{>}{\sim}}
\def \ket#1{|#1 \rangle}
\def \ob{\overline{B}^0}
\def \ol{\overline}
\begin{document}
\branch{C}   
%
\title{Theoretical Issues in $b$ Physics}
\author{J. L. Rosner \inst{1}}
\institute{Enrico Fermi Institute and Department of Physics \\
University of Chicago, 5640 S. Ellis Avenue, Chicago, IL 60637 USA \\
EFI 03-26; hep-ph/0305315; e-mail: rosner@hep.uchicago.edu}
\maketitle
\begin{abstract}
Examples are given of some current questions in $b$ physics to which LHC
experiments may provide answers.  These include (i) the precise determination
of parameters of the Cabibbo-Kobayashi-Maskawa (CKM) matrix; (ii) measurements
of CKM phases using $B$ decays to CP eigenstates; (iii) the search for direct
CP asymmetries in $B$ decays; (iv) rare radiative $B$ decays; (v) the study
of $B_s$ properties and decays, (vi) excited states of $B$ and $B_s$ mesons,
and (vii) the search for heavier quarks which could mix with the $b$ quark.
\end{abstract}
%
\section{Introduction}
The Large Hadron Collider (LHC) will permit the exploration of physics at
unprecedented energy scales toward the end of this decade, but it will also
produce $b$ quarks more copiously than any other accelerator.  If the
hadrons containing these quarks can be identified, many questions we now
face can be addressed, while undoubtedly others will arise.  In this talk I
would like to give some examples of {\it current} questions in $b$ physics to
which we would like answers.  Others may well be more timely in the LHC era.

In Section \ref{sec:CKM} we review information on weak quark transitions as
encoded in the Cabibbo-Kobayashi-Maskawa (CKM) matrix.  We then discuss CP
asymmetries in $B$ decays to CP eigenstates (Section \ref{sec:eig}) and
to self-tagging modes (``direct asymmetries,'' Section \ref{sec:dir}).
Rare radiative $B$ decays, mentioned briefly in Section \ref{sec:rad}, provide
useful information on possible new physics.  Hadron colliders such as the LHC
are the tool of choice for the study of strange $B$ ($B_s$) properties and
decays (Section \ref{sec:bs}).  Excited states of $B$ and $B_s$ mesons, for
which there have been interesting parallel developments in the charm sector,
are discussed in Section \ref{sec:exc}.  The search for heavier quarks which
which could mix with the $b$ quark is noted in Section \ref{sec:exo}, while
Section \ref{sec:sum} concludes.

\section{Weak quark transitions \label{sec:CKM}}
\begin{figure}
\includegraphics[height=3in]{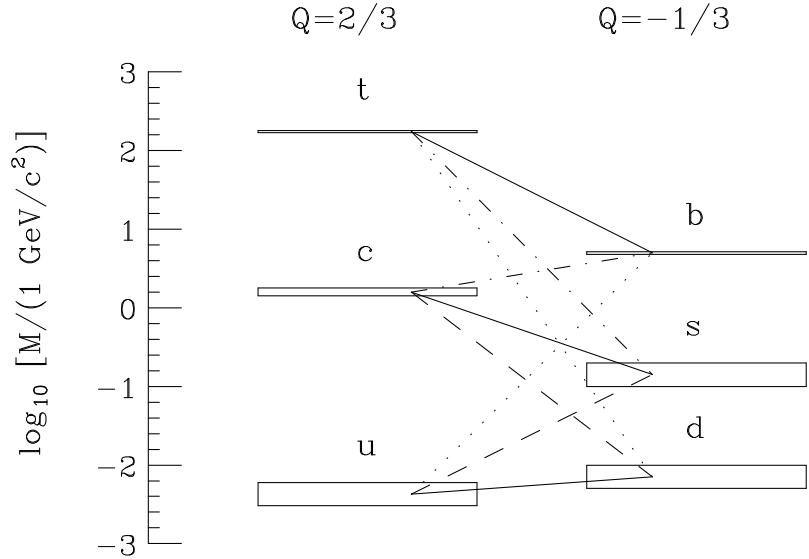}
\caption{Pattern of charge-changing weak transitions among quarks.  Solid
lines:  relative strength 1; dashed lines:  relative strength 0.22;
dot-dashed lines:  relative strength 0.04; dotted lines:  relative strength
$\le 0.01$. Breadths of horizontal lines denote estimated errors for masses.
\label{fig:trans}}
\end{figure}

The relative strengths of charge-changing weak quark transitions are
illustrated in Fig.\ \ref{fig:trans}.  This pattern is one of the central
mysteries of particle physics, along with the values of the quark masses.
We shall not address its deeper origin here, but will seek better knowledge of
strengths and phases of the transitions, to see whether all weak phenomena
including CP violation can be described satisfactorily via this pattern.

\subsection{The CKM matrix}
The interactions in Fig.\ \ref{fig:trans} may be parametrized by a unitary
$3 \times 3$ matrix, the Cabibbo-Kobayashi-Maskawa (CKM) matrix.  A convenient
form \cite{WP,Battaglia}, unitary to sufficiently high order in a small
quantity $\lambda$, is
\beq \label{eqn:WP}
V_{\rm CKM} = \left[ \begin{array}{c c c}
1 - \frac{\lambda^2}{2} & \lambda & A \lambda^3 (\rho - i \eta) \\
- \lambda & 1 - \frac{\lambda^2}{2} & A \lambda^2 \\
A \lambda^3 (1 - \bar \rho - i \bar \eta) & - A \lambda^2 & 1 \end{array}
\right]~~~,
\eeq
where $\bar \rho \equiv \rho (1 - \frac{\lambda^2}{2})$ and
$\bar \eta \equiv \eta (1 - \frac{\lambda^2}{2})$.
The columns refer to $d,s,b$ and the rows to $u,c,t$.  The parameter $\lambda =
0.224$ \cite{Battaglia} is $\sin \theta_c$, where $\theta_c$ is the Cabibbo
angle.  The value $|V_{cb}| \simeq 0.041$, obtained from $b \to c$ decays,
indicates $A \simeq 0.82$, while $|V_{ub}/V_{cb}| \simeq 0.09$, obtained from
$b \to u$ decays, implies $(\rho^2 + \eta^2)^{1/2} \simeq 0.4$.  We shall
generally use the CKM parameters quoted in Ref.\ \cite{CKMf}.

\subsection{The unitarity triangle}

\begin{figure}
\begin{center}
\includegraphics[height=1.5in]{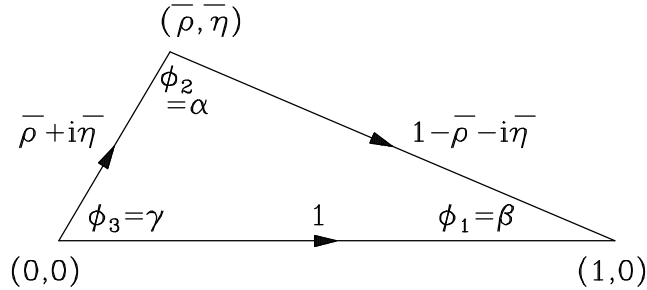}
\caption{The unitarity triangle.
\label{fig:ut}}
\end{center}
\end{figure}

The unitarity of the CKM matrix implies that the scalar product of any column
with the complex conjugate of any other column is zero; for example,
$V^*_{ub}V_{ud} + V^*_{cb} V_{cd} + V^*_{tb} V_{td} = 0$.  If one divides by
$-V^*_{cb} V_{cd}$, this relation becomes equivalent to a triangle in the
complex $\bar \rho + i \bar \eta$ plane, with vertices at (0,0) (angle $\phi_3
= \gamma$), (1,0) (angle $\phi_1 = \beta$), and $(\bar \rho, \bar \eta)$ (angle
$\phi_2 = \alpha$).  The triangle has unit base and its other two sides are
$\bar \rho + i \bar \eta = -(V^*_{ub}V_{ud}/ V^*_{cb} V_{cd})$ (opposite
$\phi_1 = \beta$) and $1 - \bar \rho - i \bar \eta = -(V^*_{tb}V_{td}/V^*_{cb}
V_{cd})$ (opposite $\phi_3 = \gamma$).  The result is shown in Fig.\
\ref{fig:ut}.

In addition to the direct measurements of CKM parameters mentioned above,
flavor-changing loop diagrams provide a number of indirect constraints.
CP-violating $K^0$--$\ol K^0$ mixing is dominated by the second-order-weak
virtual transition $\bar s d \to \bar d s$ with virtual $t \bar t$ and
$W^+ W^-$ intermediate states, and thus constrains Im$(V_{td}^2) \sim
\bar \eta (1 - \bar \rho)$, leading to a hyperbola in the $(\bar \rho,
\bar \eta)$ plane.  $B^0$--$\ol B^0$ mixing is similarly dominated by
$t \bar t$ and $W^+ W^-$ in the loop diagram for $\bar b d \to \bar d b$, and
thus constrains $|V_{td}|$ and hence $|1 - \bar \rho - i \bar \eta|$.  By
comparing $B_s$--$\ol B_s$ and $B^0$--$\ol B^0$ mixing, one can reduce
dependence on unknown matrix elements and learn a lower limit on
$|V_{ts}/V_{td}|$ or an upper limit on $|1 - \bar \rho - i \bar \eta|$.  The
range of parameters allowed at 95\% c.l.\ \cite{CKMf} is $0.08 \le \bar \rho
\le 0.34$, $0.25 \le \bar \eta \le 0.43$ (but see, e.g., \cite{Ciu} for more
a more optimistic view of our present knowledge).

\section{$B$ decays to CP eigenstates \label{sec:eig}}
One can learn CKM phases from decays of neutral $B$ mesons to CP eigenstates
$f$, where $CP \ket{f} = \xi_f \ket{f}$, $\xi_f = \pm 1$.  As a result of
$B^0$--$\ob$ mixing, a state which is $B^0$ at proper time $t=0$ will evolve
into one, denoted $B^0(t)$, which is a mixture of $B^0$ and $\ob$.  Thus
there will be one pathway to the final state $f$ from $B^0$ through the
amplitude $A$ and another from $\ob$ through the amplitude $\bar A$, which
acquires an additional phase $2 \phi_1 = 2 \beta$ through the mixing.
The interference of these two amplitudes can differ in the decays $B^0(t) \to
f$ and $\ob (t) \to f$, leading to a time-integrated rate asymmetry
\beq
{\cal A}_{CP} \equiv \frac{\Gamma(\ob \to f) - \Gamma(B^0 \to f)}
                          {\Gamma(\ob \to f) + \Gamma(B^0 \to f)}
\eeq
as well as to time-dependent rates
\beq
\left\{ \begin{array}{c} \Gamma[B^0(t) \to f] \\ \Gamma[\ob (t) \to f]
\end{array} \right\} \sim e^{- \Gamma t} [ 1 \mp {\cal A}_f \cos \Delta m t
 \mp {\cal S}_f \sin \Delta m t ]~~~,
\eeq
where
\beq
{\cal A}_f \equiv \frac{|\lambda|^2 - 1}{|\lambda|^2 + 1}~~,~~~
{\cal S}_f \equiv \frac{2 {\rm Im} \lambda}{|\lambda|^2 + 1}~~,~~~
\lambda \equiv e^{-2 i \beta} \frac{\bar A}{A}~~~,
\eeq
where ${\cal S}_f^2 + {\cal A}_f^2 \le 1$.  More details may be found in
Refs.\ \cite{BaBarPhys,TASI}.  I now note some specific cases.

\subsection{$B^0 \to J/\psi K_S$ and $\phi_1 = \beta$}
For this decay one has $\bar A/A \simeq \xi_{J/\psi K_S} = -1$.  One finds that
the time-integrated asymmetry ${\cal A}_{CP}$ is proportional to $\sin(2
\phi_1) = \sin(2 \beta)$.  Using this and related decays involving the same
$\bar b \to \bar s c \bar c$ subprocess, BaBar \cite{Babeta} finds $\sin(2
\beta) = 0.741 \pm 0.067 \pm 0.033$ while Belle \cite{Bebeta} finds $0.719 \pm
0.074 \pm 0.035$.  The two values agree with each other; the world average
\cite{avbeta} is $\sin(2 \beta) = 0.734 \pm 0.054$, consistent with other
determinations \cite{CKMf,Ciu,AL}.

\subsection{$B^0 \to \pi^+ \pi^-$ and $\phi_2 = \alpha$}
Here two amplitudes contribute to the decay:  a ``tree'' $T$ and a ``penguin''
$P$.  The decay amplitudes are
\beq
A = - (|T|e^{i \gamma} + |P| e^{i \delta})~~,~~~
\bar A = - (|T|e^{-i \gamma} + |P| e^{i \delta})~~~,
\eeq
where $\delta$ is the relative $P/T$ strong phase.  The asymmetry
${\cal A}_{CP}$ would be proportional to $\sin(2 \alpha)$ if the penguin
amplitude could be neglected.  However, one must account for its contribution.

An isospin analysis \cite{GL} of $B$ decays to $\pi^+ \pi^-$, $\pi^\pm \pi^0$,
and $\pi^0 \pi^0$ separates the contributions of decays involving $I=0$ and
$I=2$ final states.  Information can then be obtained on both strong and weak
phases.  Since the branching ratio of $B^0$ to $\pi^0 \pi^0$ may be very small,
of order $10^{-6}$, I shall discuss instead methods \cite{GR02,GRconv} in which
flavor SU(3) symmetry is used to estimate the penguin contribution
\cite{SW,GHLR,Charles}.

The tree amplitude for $B^0 (= \bar b d) \to \pi^+ \pi^-$ involves
$\bar b \to \pi^+ \bar u$, with the spectator $d$ quark combining
with $\bar u$ to form a $\pi^-$.  Its magnitude is $|T|$; its weak phase
is Arg($V^*_{ub}) = \gamma$; by convention its strong phase is 0.  The
penguin amplitude involves the flavor structure $\bar b \to \bar d$, with the
final $\bar d d$ pair fragmenting into $\pi^+ \pi^-$.  Its magnitude is
$|P|$.  The dominant $t$ contribution in the loop diagram for $\bar b \to \bar
d$ can be integrated out and the unitarity relation $V_{td} V^*_{tb} =
- V_{cd} V^*_{cb} - V_{ud} V^*_{ub}$ used.  The $V_{ud} V^*_{ub}$ contribution
can be absorbed into a redefinition of the tree amplitude, after which
the weak phase of the penguin amplitude is 0 (mod $\pi$).  By definition, its
strong phase is $\delta$.

The time-dependent asymmetries ${\cal S}_{\pi \pi}$ and ${\cal A}_{\pi \pi}$
specify both $\gamma$ (or $\alpha = \pi - \beta - \gamma$) and $\delta$,
if one has an independent estimate of $|P/T|$.  One may obtain $|P|$ from $B^+
\to K^0 \pi^+$ using flavor SU(3) \cite{SW,GHLR,GR95} and $|T|$ from $B \to
\to \pi l \nu$ using factorization \cite{LR}.  An alternative method
\cite{GRconv,Charles} uses the measured ratio of the $B^+ \to K^0 \pi^+$ and
$B^0 \to \pi^+ \pi^-$ branching ratios
to constrain $|P/T|$.  I shall discuss the first method.

In addition to ${\cal S}_{\pi \pi}$ and ${\cal A}_{\pi \pi}$, a useful quantity
is the ratio of the $B^0 \to \pi^+ \pi^-$ branching ratio $\bb(\pi^+ \pi^-)$
(averaged over $B^0$ and $\ob$) to that due to the tree amplitude alone:
\beq
R_{\pi \pi} \equiv \frac{\bb(\pi^+ \pi^-)}{\bb(\pi^+ \pi^-)|_{\rm tree}}
= 1 + 2 \left| \frac{P}{T} \right| \cos \delta \cos \gamma
+ \left| \frac{P}{T} \right|^2~~~.
\eeq
One also has
\beq
R_{\pi \pi} {\cal S}_{\pi \pi} = \sin 2 \alpha + 2 \left| \frac{P}{T} \right|
\cos \delta \sin(\beta - \alpha) - \left| \frac{P}{T} \right|^2 \sin 2 \beta
~~~,
\eeq
\beq
R_{\pi \pi} {\cal A}_{\pi \pi} = - 2 |P/T| \sin \delta \sin \gamma
~~~.
\eeq
The value of $\beta$ is specified to within a few degrees; we shall
take it to have its central value of $23.6^\circ$.  The value of $|P/T|$
(updating \cite{GR02,GRconv}) is $0.28 \pm 0.06$.  Taking the central value,
one can plot trajectories in the (${\cal S}_{\pi \pi},{\cal A}_{\pi \pi}$)
plane as $\delta$ is allowed to vary from $- \pi$ to $\pi$.

The experimental situation regarding the time-dependent asymmetries is not
yet settled.  As shown in Table \ref{tab:sa}, BaBar \cite{Bapipi} and Belle
\cite{Bepipi} obtain very different values, especially for ${\cal S}_{\pi
\pi}$.  Even if this conflict were to be resolved, however, there is a
possibility of a discrete ambiguity, since curves for different values of
$\alpha$ intersect one another.  The discrete ambiguity may be resolved with
the help of $R_{\pi \pi} = 0.62 \pm 0.28$, but the error is still too large
to be helpful.  At present values of $\phi_2 = \alpha >
90^\circ$ are favored, but with large uncertainty.  It is not yet settled
whether ${\cal A}_{\pi \pi} \ne 0$, corresponding to ``direct'' CP violation.

\begin{table}
\caption{Values of ${\cal S}_{\pi \pi}$ and ${\cal A}_{\pi \pi}$ quoted by
BaBar and Belle and their averages.  Here we have applied scale factors
$S \equiv \sqrt{\chi^2} = (2.31,1.24)$ to the errors for
${\cal S}_{\pi \pi}$ and ${\cal A}_{\pi \pi}$, respectively.
\label{tab:sa}}
\begin{center}
\begin{tabular}{c c c c} \hline \hline
    Quantity         & BaBar \cite{Bapipi}  & Belle \cite{Bepipi}            &
    Average \\ \hline
${\cal S}_{\pi \pi}$ & $0.02\pm0.34\pm0.05$ & $-1.23\pm0.41^{+0.08}_{-0.07}$ &
    $-0.49 \pm 0.61$ \\
${\cal A}_{\pi \pi}$ & $0.30\pm0.25\pm0.04$ & $ 0.77 \pm 0.27 \pm 0.08$ &
    $0.51 \pm 0.23$ \\ \hline \hline
\end{tabular}
\end{center}
\end{table}

\subsection{$B^0 \to \phi K_S$ vs. $B^0 \to J/\psi K_S$}
In $B^0 \to \phi K_S$, governed by the $\bar b \to \bar s$ penguin amplitude,
the standard model predicts the same CP asymmetries as in those processes (like
$B^0 \to J/\psi K_S$)  governed by $\bar b \to \bar s c \bar c$.  In both cases
the weak phase is expected to be 0 (mod $\pi$), so the indirect CP asymmetry
should be governed by $B^0$--$\ob$ mixing and thus should be proportional to
$\sin 2 \beta$.  There should be no direct CP asymmetries (i.e., ${\cal A}
\simeq 0$) in either case.  This is true for $B \to J/\psi K$; ${\cal A}$ is
consistent with zero in the neutral mode, while the direct CP asymmetry is
consistent with zero in the charged mode \cite{Babeta}.  However, a different
result for $B^0 \to \phi K_S$ could point to new physics in the $\bar b \to
\bar s$ penguin amplitude \cite{GW}.

The experimental asymmetries in $B^0 \to \phi K_S$ \cite{Baphks,Bephks} are
shown in Table \ref{tab:phks}. For ${\cal A}_{\phi K_S}$ there is a substantial
discrepancy between BaBar and Belle.  The value of ${\cal S}_{\phi K_S}$, which
should equal $\sin 2 \beta = 0.734 \pm 0.054$ in the standard model, is about
$2.7 \sigma$ away from it.  If the amplitudes for $B^0 \to \phi K^0$ and $B^+
\to \phi K^+$ are equal (true in many approaches), the time-integrated CP
asymmetry $A_{CP}$ in the charged mode should equal ${\cal A}_{\phi K_S}$.  The
BaBar Collaboration \cite{Aubert:2003tk} has recently reported
$A_{CP} = 0.039 \pm 0.086 \pm 0.011$.

\begin{table}
\caption{Values of ${\cal S}_{\phi K_S}$ and ${\cal A}_{\phi K_S}$ quoted by
BaBar and Belle and their averages.  Here we have applied a scale factor of
$\sqrt{\chi^2} = 2.29$ to the error on ${\cal A}_{\phi K_S}$.
\label{tab:phks}}
\begin{center}
\begin{tabular}{c c c c} \hline \hline
    Quantity         & BaBar \cite{Baphks}  & Belle \cite{Bephks}            &
    Average \\ \hline
${\cal S}_{\phi K_S}$ & $-0.18\pm0.51\pm0.07$ & $-0.73\pm0.64\pm0.22$ &
 $-0.38 \pm 0.41$ \\
${\cal A}_{\phi K_S}$ & $0.80\pm0.38\pm0.12$ & $-0.56\pm0.41\pm0.16$ &
 $0.19 \pm 0.68$ \\ \hline \hline
\end{tabular}
\end{center}
\end{table}

Many proposals for new physics can account for the departure of ${\cal S}_
{\phi K_S}$ from its expected value of $\sin 2 \beta$
\cite{npphks}.  A method similar to that \cite{GR02,GRconv} used in analyzing
$B^0 \to \pi \pi$ for extracting a new physics amplitude has been developed
in collaboration with Cheng-Wei Chiang \cite{CR03}.  One uses the measured
values of ${\cal S}_{\phi K_S}$ and ${\cal A}_{\phi K_S}$ and the ratio
\beq \label{eqn:rphks}
R_{\phi K_S} \equiv \frac{\bb(B^0 \to \phi K_S)}{\bb(B^0 \to \phi K_S)|_{\rm
 std}} = 1 + 2 r \cos \phi \cos \delta + r^2~~~,
\eeq
where $r$ is the ratio of the magnitude of the new amplitude to the one in
the standard model, and $\phi$ and $\delta$ are their relative weak and
strong phases.  For any values of $R_{\phi K_S}$, $\phi$, and $\delta$, Eq.\
(\ref{eqn:rphks}) can be solved for the amplitude ratio $r$ and one then
calculates
\bea
R_{\phi K_S} {\cal S}_{\phi K_S} & = & \sin 2 \beta + 2 r \cos \delta
\sin(2 \beta - \phi) + r^2 \sin 2(\beta - \phi)~~\\
R_{\phi K_S} {\cal A}_{\phi K_S} & = & 2 r \sin \phi \sin \delta~~~.
\eea
The $\phi K_S$ branching ratio in the standard model is calculated using the
penguin amplitude from $B^+ \to K^{*0} \pi^+$ and an estimate of electroweak
penguin corrections.  It was found \cite{CR03} that $R_{\phi K_S} = 1.0 \pm
0.2$.

Various regions of $(\phi, \delta)$ can reproduce the observed values of
${\cal S}_{\phi K_S}$ and ${\cal A}_{\phi K_S}$.  As errors on the observables
shrink, so will the allowed regions.  However, there will
always be a solution for {\it some} $\phi$ and $\delta$ as long as $R$
remains compatible with 1.  (The allowed regions of $\phi$ and $\delta$ are
restricted if $R \ne 1$ \cite{CR03}.)  Typical values of $r$ are of order 1;
one generally needs to invoke new-physics amplitudes comparable to those in
the standard model.

The above scenario envisions new physics entirely in $B^0 \to \phi K^0$ and
not in $B^+ \to K^{*0} \pi^+$.  An alternative is that new physics
contributes to the $\bar b \to \bar s$ penguin amplitude and thus appears
in {\it both} decays.  Here it is convenient to define a ratio
\beq
R' \equiv \frac{\gb(B^0 \to \phi K^0)}{\gb(B^+ \to K^{*0} \pi^+)}~~~,
\eeq
where $\gb$ denotes a partial width averaged over a process and its CP
conjugate.  Present data indicate $R' = 0.78 \pm 0.17$.  The $B^0 \to \phi K^0$
amplitude contains a contribution from both the gluonic and electroweak penguin
terms, while $B^+ \to K^{*0} \pi^+$ contains only the former.  Any departure
from the expected ratio of the electroweak to gluonic penguin amplitudes
would signify new physics.  Again, the central value of ${\cal S}$ would
suggest this to be the case \cite{CR03}, but one must wait until the
discrepancy with the standard model becomes more significant.  At present
both the decays $B^0 \to K_S (K^+ K^-)_{CP = +}$ and $B^0 \to \eta' K_S$
display CP asymmetries consistent with standard expectations.

\subsection{$B^0 \to K_S (K^+ K^-)_{CP=+}$}
The Belle Collaboration \cite{Bephks} finds that for $K^+ K^-$ not in the
$\phi$ peak, most of the decay $B^0 \to K_S K^+ K^-$ involves even CP for the
$K^+ K^-$ system ($\xi_{K^+ K^-} = +1$).  It is found that
\bea
- \xi_{K^+ K^-} {\cal S}_{K^+ K^-} & = & 0.49\pm0.43\pm 0.11^{+0.33}_{-0.00}
~~~,\\
{\cal A}_{K^+ K^-} & = & -0.40 \pm 0.33 \pm 0.10^{+0.00}_{-0.26}~~,
\eea
where the third set of errors arise from uncertainty in the fraction of the
CP-odd component.  Independent estimates of this fraction have been performed
in Refs.\ \cite{GLNQ} and \cite{GRKKK}.  The quantity $- \xi_{K^+ K^-} {\cal
S}_{K^+ K^-}$ should equal $\sin 2 \beta$ in the standard model, but additional
non-penguin contributions can lead this quantity to range between 0.2 and
1.0 \cite{GRKKK}.

\subsection{$B \to \eta' K$ (charged and neutral modes)}
At present neither the rate nor the CP asymmetry in $B \to \eta' K$ present
a significant challenge to the standard model.  The rate can be reproduced
with the help of a modest contribution from a ``flavor-singlet penguin''
amplitude, the need for which was pointed out \cite{DGR95} prior to the
observation of this decay.  One only needs to boost the standard penguin
amplitude's contribution by about 50\% via the flavor-singlet term in order to
explain the observed rate \cite{DGR97,CR01,FHH,CGR}.  (An alternative treatment
\cite{BN} finds an enhanced standard-penguin contribution to $B \to \eta' K$.)
The CP asymmetry is
not a problem; the ordinary and singlet penguin amplitudes are expected
to have the same weak phase Arg$(V^*_{ts}V_{tb}) \simeq \pi$ and hence one
expects ${\cal S}_{\eta' K_S} \simeq \sin 2 \beta$, ${\cal A}_{\eta' K_S}
\simeq 0$.  The experimental situation is shown in Table \ref{tab:etapks}.
The value of ${\cal S}_{\eta' K_S}$ is consistent with the standard model
expectation at the $1 \sigma$ level, while ${\cal A}_{\eta' K_S}$ is consistent
with zero.

\begin{table}
\caption{Values of ${\cal S}_{\eta' K_S}$ and ${\cal A}_{\eta' K_S}$ quoted by
BaBar and Belle and their averages.  Here we have applied scale factors
$S \equiv \sqrt{\chi^2} = (1.48,1.15)$ to the errors for
${\cal S}_{\eta' K_S}$ and ${\cal A}_{\eta' K_S}$, respectively.
\label{tab:etapks}}
\begin{center}
\begin{tabular}{c c c c} \hline \hline
    Quantity         & BaBar \cite{Baphks}  & Belle \cite{Bephks}            &
    Average \\ \hline
${\cal S}_{\eta' K_S}$ & $0.02\pm0.34\pm0.03$ & $0.76\pm0.36^{+0.05}_{-0.06}$ &
$0.37 \pm 0.37$  \\
${\cal A}_{\eta' K_S}$ & $-0.10\pm0.22\pm0.03$ & $0.26\pm0.22\pm0.03$ &
$0.08 \pm 0.18$  \\ \hline \hline
\end{tabular}
\end{center}
\end{table}

The singlet penguin amplitude may contribute elsewhere in $B$ decays.  It is
a possible source of a low-effective-mass $\bar p p$ enhancement \cite{Kpp} in
$B^+ \to \bar p p K^+$ \cite{JRbbbar}.

\section{Direct CP asymmetries \label{sec:dir}}
Decays such as $B \to K \pi$ (with the exception of $B^0 \to K^0 \pi^0$) are
{\it self-tagging}, i.e., their final states indicate the flavor of the
decaying state.  For example, the $K^+ \pi^-$ final state is expected to
originate purely from a $B^0$ and not from a $\ob$.  Since such self-tagging
decays do not involve a CP eigenstate, they involve both weak and strong
phases.  Several methods permit one to separate these from one another.  We
give some examples.

\subsection{$B^0 \to K^+ \pi^-$ vs.\ $B^+ \to K^0 \pi^+$}
The decay $B^+ \to K^0 \pi^+$ is a pure penguin ($P$) process, while the
amplitude for $B^0 \to K^+ \pi^-$ is proportional to $P + T$, where $T$ is a
(strangeness-changing) tree amplitude.  The ratio $T/P$ has magnitude $r$, weak
phase $\gamma \pm \pi$, and strong phase $\delta$.  The ratio
$R_0$ of these two rates (averaged over a process and its CP conjugate) is
\beq \label{eqn:Rval}
R_0 \equiv \frac{\gb(B^0 \to K^+ \pi^-)}{\gb(B^+ \to K^0 \pi^+)} =
1 - 2 r \cos \gamma \cos \delta + r^2 \ge \sin^2 \gamma~~~,
\eeq
where the inequality holds for any $r$ and $\delta$.  For $R_0 < 1$
this inequality can be used to impose a useful constraint on
$\gamma$ \cite{FM}.  On the basis of branching ratios \cite{Babrs,Bebrs,CLbrs}
summarized in Ref.\ \cite{JRmor} and using the $B^+/B^0$ lifetime ratio from
Ref.\ \cite{LEPBOSC}, one finds $R_0 = 0.99 \pm 0.09$, which is
consistent with 1 and does not permit application of the bound.  However,
using additional information on $r$ and the CP asymmetry in $B^0 \to K^+
\pi^-$, one can obtain a constraint on $\gamma$ \cite{GR02,GRKpi}.

The CP asymmetry ${\cal A}_{CP}$ (2) can be written for $B^0 \to K^+ \pi^-$ as
\beq \label{eqn:asy}
{\cal A}_{CP}(K^+ \pi^-) \equiv \frac{\Gamma(\ob \to K^- \pi^+)
- \Gamma(B^0 \to K^+ \pi^-)}{2 \gb(B^0 \to K^+ \pi^-)} = - \frac{2 r \sin
\gamma \sin \delta}{R_0}~~~.
\eeq

\begin{figure}
\begin{center}
\includegraphics[height=3.5in]{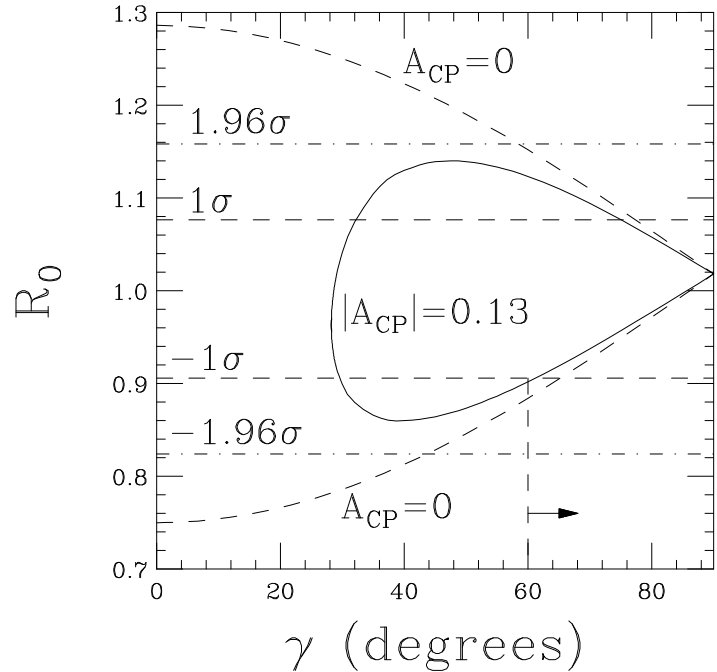}
\caption{Behavior of $R_0$ for $r = 0.13$ and ${\cal A}_{CP}(K^+ \pi^-) = 0$
(dashed curves) or $|{\cal A}_{CP}(K^+ \pi^-)| = 0.13$ (solid curve) as a
function of the weak phase $\gamma$.
Horizontal dashed lines denote $\pm 1 \sigma$ experimental limits on $R_0$,
while dot-dashed lines denote $95\%$ c.l. ($\pm 1.96 \sigma$) limits.
The upper branches of the curves correspond to the case $\cos \gamma
\cos \delta <0$, while the lower branches correspond to $\cos \gamma
\cos \delta >0$.
\label{fig:R0}}
\end{center}
\end{figure}

One may eliminate $\delta$ between this equation and Eq.\ (\ref{eqn:Rval})
and plot $R_0$ as a function of $\gamma$ for the allowed range of
${\cal A}_{CP}(K^+ \pi^-)$.  The value of $r$, based on present branching
and arguments given in Refs.\ \cite{GR02,JRmor,GRKpi}), is $r=0.17 \pm 0.04$.
The latest BaBar and Belle data imply ${\cal A}_{CP}(K^+ \pi^-) = -0.09
\pm 0.04$ \cite{CGR}, leading us to take $|{\cal A}_{CP}(K^+ \pi^-)| \le 0.13$
at the $1 \sigma$ level.  Curves for ${\cal A}_{CP} =0$ and $|{\cal A}_{CP}| =
0.13$ (the $K^+ \pi^-$ final state is to be understood) are
shown in Fig.\ \ref{fig:R0}. The lower limit $r = 0.13$ is used to generate
these curves since the limit on $\gamma$ will be the most conservative.

At the $1 \sigma$ level, using the constraints that $R_0$ must lie between 0.90
and 1.08 and $|{\cal A}_{CP}|$ must lie between zero and 0.13, one can
establish that
$\gamma \gs 60^\circ$.  No bound can be obtained at the 95\% confidence level,
however.  Despite the impressive improvement in experimental precision (a
factor of 2 decrease in errors since the analysis of Ref.\ \cite{GR02}),
further data are needed in order for a useful constraint to be obtained.

\subsection{$B^+ \to K^+ \pi^0$ vs.\ $B^+ \to K^0 \pi^+$}
The comparison of rates for $B^+ \to K^+ \pi^0$ and $B^+ \to K^0 \pi^+$ also
can give information on $\gamma$.  The amplitude for $B^+ \to K^+ \pi^0$ is
proportional to $P + T + C$, where $C$ is a color-suppressed amplitude.
Originally it was suggested that this amplitude be compared with $P$ from $B^+
\to K^0 \pi^+$ and $T+C$ taken from $B^+ \to \pi^+ \pi^0$ using flavor SU(3)
\cite{GRL} using a triangle construction to determine $\gamma$.  However,
electroweak penguin amplitudes contribute significantly in the $T+C$ term
\cite{EWP}.  It was noted subsequently \cite{NR} that since the $T+C$ amplitude
corresponds to isospin $I(K \pi) = 3/2$ for the final state, the
strong-interaction phase of its EWP contribution is the same as that of the
rest of the $T+C$ amplitude, permitting the calculation of the EWP correction.

\begin{figure}
\begin{center}
\includegraphics[height=3.5in]{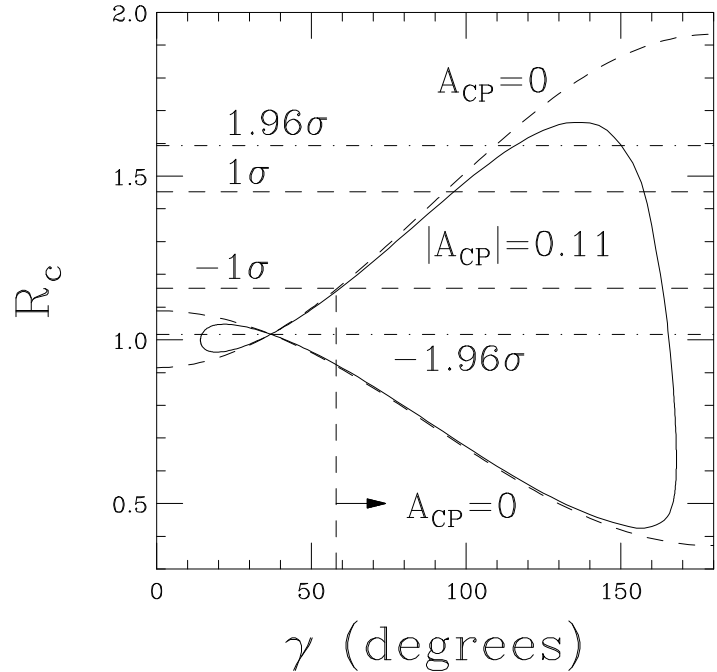}
\caption{Behavior of $R_c$ for $r_c = 0.22$ ($1 \sigma$ upper limit) and
${\cal A}_{CP}(K^+ \pi^0) = 0$ (dashed curves) or $|{\cal A}_{CP}(K^+ \pi^0)|
= 0.11$ (solid curve) as a function of the
weak phase $\gamma$. Horizontal dashed lines denote $\pm 1 \sigma$ experimental
limits on $R_c$, while dotdashed lines denote 95\% c.l. ($ \pm 1.96 \sigma$)
limits.  Upper branches of curves correspond to $\cos \delta_c(\cos \gamma -
\delta_{EW}) < 0$, while lower branches
correspond to $\cos \delta_c(\cos \gamma - \delta_{EW}) > 0$.  Here we have
taken $\delta_{EW} = 0.80$ (its $1 \sigma$ upper limit), which
leads to the most conservative bound on $\gamma$.
\label{fig:Rc}}
\end{center}
\end{figure}

New data on branching ratios and CP asymmetries permit an update of previous
analyses \cite{GR02,NR}.  One makes use of the quantities (see \cite{CGR} for
values)
\bea
R_c & \equiv & \frac{2 \gb(B^+ \to K^+ \pi^0)}{\gb(B^+ \to K^0 \pi^+)}
= 1 - 2 r_c \cos \delta_c~(\cos \gamma - \delta_{\rm EW}) \nonumber \\
& + & r_c^2(1 - 2 \delta_{EW} \cos \gamma + \delta_{EW}^2) 
= 1.30 \pm 0.15~~, \label{eqn:Rc}
\eea
\beq \label{eqn:Accp}
{\cal A}_{CP}(K^+ \pi^0) 
 =  - \frac{2 r_c \sin \delta_c \sin \gamma}{R_c} = 0.035 \pm 0.071~~~,
\eeq
where $r_c \equiv |(T+C)/P| = 0.20 \pm 0.02$, and $\delta_c$ is a strong
phase, eliminated by combining (\ref{eqn:Rc}) and (\ref{eqn:Accp}).
One must also use an estimate \cite{NR} of the electroweak penguin parameter
$\delta_{\rm EW} = 0.65 \pm 0.15$.  One obtains the most conservative (i.e.,
weakest) bound on $\gamma$ for the maximum values of $r_c$ and $\delta_{\rm
EW}$ \cite{GR02}.  The resulting plot is shown in Fig.\ \ref{fig:Rc}.  One
obtains a bound at the $1 \sigma$ level very similar to that in the previous
case:  $\gamma \gs 58^\circ$.  The bound is actually set by the curve for
{\it zero} CP asymmetry, as emphasized in Ref.\ \cite{NR}.

\subsection{$B^+ \to \pi^+ \eta$}
The possibility that several different amplitudes could contribute to
$B^+ \to \pi^+ \eta$, thereby leading to the possibility of a large direct
CP asymmetry, has been recognized for some time \cite{GR95,DGR95,DGR97,BRS,AK}.
Contributions can arise from a tree amplitude (color-favored plus
color-suppressed) $T+C$, whose magnitude is estimated to be $\sqrt{2/3}$ that
occurring in $B^+ \to \pi^+ \pi^0$, a penguin amplitude $P$, obtained via
flavor SU(3) from $B^+ \to K^0 \pi^+$, and a singlet penguin amplitude $S$,
obtained from $B \to \eta' K$.

In Table \ref{tab:etapi} we summarize branching ratios and CP asymmetries
obtained for the decay $B^+ \to \pi^+ \eta$ by CLEO \cite{CLeta}, BaBar
\cite{Baeta}, and Belle \cite{Bebrs}.  We assume that the $S$ and $P$
amplitudes have the same weak and strong phases.  The equality of their weak
phases is quite likely, while tests exist for the latter assumption \cite{CGR}.

\begin{table}
\caption{Branching ratios and CP asymmetries for $B^ \to \pi^+ \eta$.
\label{tab:etapi}}
\begin{center}
\begin{tabular}{l c c} \hline \hline
 & $\bb~(10^{-6})$ & $A_{CP}$ \\ \hline
CLEO \cite{CLeta}  & $1.2^{+2.8}_{-1.2}~(< 5.7)$ & -- \\
BaBar \cite{Baeta} & $4.2^{+1.0}_{-0.9} \pm 0.3$ & $-0.51^{+0.20}_{-0.18}$ \\
Belle \cite{Bebrs} & $5.2^{+2.0}_{-1.7} \pm 0.6$ & --  \\
Average            & $4.1 \pm 0.9$               & $-0.51^{+0.20}_{-0.18}$ \\
$|T+C|^2$ alone    &             3.5             &          0             \\
$|P+S|^2$ alone    &             1.9             &          0        \\
\hline \hline
\end{tabular}
\end{center}
\end{table}

If the amplitude $A$ for a process receives two contributions with differing
strong and weak phases, one can write
\beq
A = a_1 + a_2 e^{i \phi} e^{i \delta}~~,~~~
\bar A = a_1 + a_2 e^{-i \phi} e^{i \delta}~~~.
\eeq
The CP-averaged decay rate is proportional to $a_1^2 + a_2^2 + 2 a_1 a_2
\cos \phi \cos \delta$, while the CP asymmetry is
\beq
A_{CP} = - \frac{2 a_1 a_2 \sin \phi \sin \delta}
{a_1^2 + a_2^2 + 2 a_1 a_2 \cos \phi \cos \delta}~~~.
\eeq
In the case of $B^+ \to \pi^+ \eta$ the rates and CP asymmetry suggest that
$|\sin \phi \sin \delta| > |\cos \phi \cos \delta|$.  Details of this pattern
and its implications for other processes are described in Ref.\ \cite{CGR}.
It is predicted there that $\bb(B^+ \to \pi^+ \eta') =
(2.7 \pm 0.7) \times 10^{-6}$ (below current upper bounds) and that
$A_{CP}(\pi^+ \eta') = -0.57 \pm 0.23$.

\section{Rare radiative $B$ decays \label{sec:rad}}
A number of processes in which a $B$ or $B_s$ decays to final states with
photons or lepton pairs are particularly sensitive to non-standard physics.
An example is $B_s \to \mu^+ \mu^-$, for which the standard model predicts
$\b(B_s \to \mu^+ \mu^-) = (3.1 \pm 1.4) \times 10^{-9}$ \cite{Bobeth}.
Charged Higgs boson exchanges or other effects could enhance this branching
ratio significantly while respecting the constraint associated with the
branching ratio for $b \to s \gamma$, which appears compatible with standard
model predictions.  For a good discussion of this process and of $B \to X_s
\ell^+ \ell^-$ see Ref.\ \cite{Hiller}, as well as several presentations at the
present conference \cite{CMSAT}.  In the latter decay the forward-backward
asymmetries
exhibit interesting behavior as a function of $m(\ell^+ \ell^-)$, with signs
and a characteristic zero in the standard model which can be different in
variant theories.

\section{$B_s$ properties and decays \label{sec:bs}}
\subsection{$B_s$--$\overline B_s$ mixing}
The ratio of the $B_s$--$\overline B_s$ mixing amplitude $\Delta m_s$ to the
$B^0$--$\ol B^0$ mixing amplitude $\Delta m_d$ ($B_d \equiv B^0$) is given by
\beq
\frac{\Delta m_s}{\Delta m_d} =
\frac{f_{B_s}^2 B_{B_s}}{f_{B_d}^2 B_{B_d}} \frac{m_{B_s}}{m_{B_d}}
\left| \frac{V_{ts}}{V_{td}} \right|^2 \simeq 48 \times 2^{\pm 1}~~~.
\eeq
Here $f_{B_{d,s}}$ are meson decay constants, while $B_{B_{d,s}}$ are
numbers of order 1 expressing the degree to which the mixing amplitude can be
calculated by saturating with vacuum intermediate states.  The latest lattice
estimate of the ratio $\xi \equiv (f_{B_s}/f_{B_d})\sqrt{B_{B_s}/B_{B_d}}$
is $1.21 \pm 0.04 \pm 0.05$ \cite{Bec}.  We have taken a generous range
\beq
|V_{td}| = A \lambda^3|1 - \bar \rho - i \bar \eta| = (0.8 \pm 0.2) A \lambda^3
\eeq
with $|V_{ts}| = A \lambda^2$ and $\lambda = 0.22$.  With \cite{LEPBOSC}
$\Delta m_d = 0.503 \pm 0.007$ ps$^{-1}$ one then predicts
\beq
\Delta m_s = 24~{\rm ps}^{-1} \times 2^{\pm 1}~~~.
\eeq
The lower portion of this range is already excluded by the bound \cite{LEPBOSC}
\beq
\Delta m_s > 14.4~{\rm ps}^{-1}~(95\%~{\rm c.l.})~~~.
\eeq
When $\Delta m_s$ is measured it is likely to be known fairly well
immediately, and will constrain $\bar \rho$ significantly.

\subsection{Decays to CP eigenstates}
\subsubsection{$B_s \to J/\psi \phi,~J/\psi \eta, \ldots$.}
Since the weak phase in $\bar b \to \bar c c \bar s$ is expected to be zero
while that of $B_s$--$\ol B_s$ mixing is expected to be very small [in the
parametrization of Eq.\ (1) an imaginary part Im($V_{ts})= -A \lambda^4 \eta$
was not
written explicitly], one expects CP asymmetries to be only a few percent in
the standard model for those $B_s$ decays dominated by this quark subprocess.
The $B_s \to J/\psi \phi$ final state is not a CP eigenstate but the even and
odd CP components can be separated using an angular analysis.  The final
states of $B_s \to J/\psi \eta$ and $B_s \to J/\psi \eta'$ are CP-even so no
such analysis is needed.

\subsubsection{$B_s \to K^+ K^-$ vs.\ $B^0 \to \pi^+ \pi^-$.}
A comparison of time-dependent asymmetries in $B_s \to K^+ K^-$ and $B^0 \to
\pi^+ \pi^-$ \cite{RFKK} allows one to separate out strong and weak phases
and relative tree and penguin contributions.  In $B_s \to K^+ K^-$ the $\bar b
\to \bar s$ penguin amplitude is dominant, while the strangeness-changing
tree amplitude $\bar b \to \bar u u \bar s$ is subsidiary.  In $B^0 \to \pi^+
\pi^-$ it is the other way around: The $\bar b \to \bar u u \bar d$ tree 
amplitude dominates, while the $\bar b \to \bar d$ penguin is
Cabibbo-suppressed.  The U-spin subgroup of SU(3), which interchanges $s$ and
$d$ quarks, relates each amplitude in one process to that in the other aside
from the CKM factors.

\subsubsection{$\overline B_s,~B^0 \to K^+ \pi^-$.}
A potential problem with $B_s \to K^+ K^-$ and $B^0 \to \pi^+ \pi^-$ is that
the mass peaks will overlap with one another if analyzed in terms of the same
final state (e.g., $\pi^+ \pi^-$) \cite{Jesik}.  Thus, in the absence of good
particle identification, a variant on this scheme employing the decays
$B^0 \to K^+ \pi^-$ and $B_s \to K^- \pi^+$ (also related to one another by
U-spin) may be useful \cite{GRKpi00}.  For these final states, kinematic
separation may be easier.  A further variant is to study the time-dependence
of $B_s \to K^+ K^-$ while normalizing the penguin amplitude using $B_s \to
K^0 \ol K^0$ \cite{GRKK}.

\subsection{Other SU(3) relations}
The U-spin subgroup of SU(3) allows one to relate many other $B_s$ decays
besides those mentioned above to corresponding $B_d$ decays \cite{MGU}.
Particularly useful are relations between CP-violating rate {\it differences}.
One thus will have the opportunity to perform many tests of flavor SU(3) and
to learn a great deal more about final-state phase patterns when a variety of
$B_s$ decays can be studied.

\section{Excited states \label{sec:exc}}
\subsection{Flavor tagging for neutral $B$ mesons}
One promising method for tagging the flavor of a neutral $B$ meson is to
study the charge of the leading light hadron accompanying the fragmentation
of the heavy quark.  This method was initially proposed by Ali and Barreiro
\cite{AB} to identify the flavor of a $B_s$ via the charge of the accompanying
kaon.  It was utilized in Refs.\ \cite{GNR,GRtag} to distinguish $B^0$'s from
$\ol B^0$'s.  An initial $b$ will fragment into a $\ol B^0$ by ``dressing''
itself with a $\bar d$.  The accompanying $d$, if incorporated into a charged
pion, will end up in a $\pi^-$.  Thus a $\pi^-$ is more likely to be ``near'' a
$\ol B^0$ than to a $B^0$ in phase space.  This correlation
between $\pi^-$ and $\ol B^0$ (and the corresponding correlation between
$\pi^+$ and $B^0$) is also what one would expect on the basis of non-exotic
resonance formation.  Thus the study of the resonance spectrum of the excited
$B$ mesons which can decay to $B + \pi$ or $B^* + \pi$ is of special
interest \cite{EHQ}.  The lowest such mesons are the P-wave levels of a $\bar
b$ antiquark and a light ($u$ or $d$) quark.

\subsection{Surprise:  Excited $D_s$ state below $DK$ threshold}
A new sensation has been reported by the BaBar Collaboration \cite{BaDs}
and confirmed by CLEO \cite{CLDs}.  Partial information on the P-wave levels of
a charmed quark $c$ and an antistrange $\bar s$ consists of candidates for
$J=1$ and $J=2$ states at 2535 and 2572 MeV \cite{PDG}.  These levels have
narrow widths and are behaving as would be expected if the spin of the $\bar s$
and the orbital angular momentum were coupled up to $j = 3/2$.  (One expects
$j$-$j$ rather than $L$-$S$ coupling in
a light-heavy system \cite{DGG,JRPW,HQ}.)  If the $j=1/2$
states were fairly close to these in mass one would then expect another $J=1$
state and a $J=0$ state somewhere above 2500 MeV.  Instead, a candidate for a
$J=0$ $c \bar s$ state has been found around 2317 MeV, with the second $J=1$
level around 2463 MeV.  Both are narrow, since they are too light to decay
respectively to $D K$ or $D^* K$.  They decay instead via the isospin-violating
transitions $D_{s0}(2317) \to D_s \pi^0$ and $D_{s1}(2463) \to D_s^* \pi^0$.
They are either candidates for $D^{(*)} K$ molecules \cite{BCL}, or indications
of a broken chiral symmetry which places them as positive-parity partners of
the $D_s$ and $D_s^*$ negative-parity $c \bar s$ ground states \cite{BEH}.
Indeed, the mass splittings between the parity partners appear to be exactly as
predicted ten years ago \cite{BH}.  Potential-based quarkonium models have a
hard time accommodating such low masses \cite{CJ,SG,Col},

There should exist {\it non-strange} $j=1/2$ $0^+$ and $1^+$ states, lower in
mass than the $j=3/2$ states at 2422 and 2459 MeV \cite{PDG} but quite broad
since their respective $\ol B \pi$ and $\ol B^* \pi$ channels will be open.
The study of such states will be of great interest since the properties of the
corresponding $B$-flavored states will be useful in tagging the flavor of
neutral $B$ mesons, as noted in the previous subsection.

\subsection{Narrow positive-parity states below $\ol B^{(*)} K$ threshold?}
If a strange antiquark can bind to a charmed quark in both negative- and
positive-parity states, the same must be true for a strange antiquark and
a $b$ quark.  One should then expect to see narrow $J^P = 0^+$ and $1^+$
states with the quantum numbers of $\ol B K$ and $\ol B^* K$ but below those
respective thresholds.  They should decay to $\ol B_s \pi^0$ and $\ol B_s^*
\pi^0$, respectively.  To see such decays one will need a multi-purpose
detector with good charged particle and $\pi^0$ identification!  Such
detectors are envisioned for both the Tevatron \cite{BTeV} and the LHC
\cite{LHCb}.

\section{Exotic $Q=-1/3$ quarks \label{sec:exo}}
Might there be heavier quarks visible at hadron colliders?  At present we
have evidence for three families of quarks and leptons belonging to
16-dimensional multiplets of the grand unified group SO(10) (counting
right-handed neutrinos as a reasonable explanation of the observed oscillations
between different flavors of neutrinos).  Now, just as SO(10) was pieced
together from multiplets of SU(5) with dimensions 1, 5, and 10, we can imagine
a still larger grand unified group whose smallest representation contains the
16-dimensional SO(10) spinor.  Such a group is the exceptional group E$_{\rm
6}$ \cite{GRS}.  Its smallest representation, of dimension 27, contains a
16-dimensional spinor, a 10-dimensional vector, and a singlet of SO(10).  The
10-dimensional vector contains vector-like isosinglet quarks ``$h$'' and
antiquarks $\bar h$ of charge
$Q = \pm 1/3$ and isodoublet leptons.  The SO(10) singlets are candidates for
sterile neutrinos, one for each family.

The new exotic $h$ quarks can mix with the $b$ quark and push its mass
down with respect to the top quark \cite{JRmix}.  Troy Andre and I are
currently looking at signatures of $h \bar h$
production in hadron colliders, with an eye to either setting lower mass
limits or seeing such quarks through their decays to $Z + b$, $W + t$, and
possibly ${\rm Higgs} + b$.  The $Z$, for example, would be identified by its
decays to $\nu \bar \nu$, $\ell^+ \ell^-$, or jet $+$ jet, while the Higgs
boson would show up through its $b \bar b$ decay if it were far enough below
$W^+ W^-$ threshold.

\section{Summary \label{sec:sum}}
The process $B^0 \to J/\psi K_S$ has provided spectacular confirmation of the
Kobayashi-Maskawa theory of CP violation, measuring $\beta$ to a few
degrees.  Now one is entering the territory of more difficult measurements.

The decay $B^0 \to \pi^+ \pi^-$ has great potential for giving useful
information on $\alpha$.  One needs either a measurement of
${\cal B}(B^0 \to \pi^0 \pi^0)$ \cite{GL}, probably at the $10^{-6}$ level
(present limits \cite{Babrs,Bebrs,CLbrs} are several times that), or a
better estimate of the tree amplitude from $B \to \pi l \nu$ \cite{LR}.
Indeed, such an estimate has been presented recently \cite{LR03}.  The
BaBar and Belle experimental CP asymmetries \cite{Bapipi,Bepipi} will
eventually converge to one another, as did the initial measurements
of $\sin 2 \beta$ using $B^0 \to J/\psi K_S$.

The $B \to \phi K_S$ decay can display new physics via special $\bar b \to \bar
s s \bar s$ operators or effects on the $\bar b \to \bar s$ penguin.  Some
features of any new amplitude can be extracted from the data in a
model-independent way if one uses both rate and asymmetry information
\cite{CR03}.  While the effective value of $\sin 2 \beta$ in $B^0 \to \phi K_S$
seems to differ from its expected value by more than $2 \sigma$, CP asymmetries
in $B \to K_S (K^+ K^-)_{CP=+}$ do not seem anomalous.

The rate for $B \to \eta' K_S$ is not a problem for the standard model if one
allows for a modest flavor-singlet penguin contribution in addition to the
standard penguin amplitude.  The CP asymmetries for this process are in accord
with the expectations of the standard model at the $1 \sigma$ level or
better.  Effects of the singlet penguin amplitude may also be visible
elsewhere, for example in $B^+ \to p \bar p K^+$.

Various ratios of $B \to K \pi$ rates, when combined with information on
CP asymmetries, show promise for constraining phases in the CKM matrix.
These tests have shown a steady improvement in accuracy since the asymmetric
$B$ factories have been operating.  One expects further progress as 
instantaneous and accumulated $e^+ e^-$ luminosities increase, and as hadron
colliders begin to provide important contributions.  The decays $B^+ \to
\pi^+ \eta$ and $B^+ \to \pi^+ \eta'$ show promise for displaying large CP
asymmetries \cite{CGR} since they involve contributions of different amplitudes
with comparable magnitudes.

Rare decays of nonstrange and strange $B$'s involving photons or lepton pairs
are beginning to be studied in detail, and the LHC will be able to look for
the rare and interesting $B_s \to \mu^+ \mu^-$ decay which can greatly exceed
its standard model value in some theories.  In the near term
the prospects for learning about the $B_s$--$\ol B_s$ mixing amplitude are
good.  One hopes that this will be an early prize of Run II at the Tevatron.
The study of CP violation and branching ratios in $B_s$ decays will be an
almost exclusive province of hadron colliders, whose potentialities will be
limited only by the versatility of detectors.  Surprises in spectroscopy,
as illustrated by the low-lying positive-parity $c \bar s$ candidiates, still
can occur, and one is sure to find more surprises at the Tevatron and the LHC.
Finally, one can search for objects related to the properties of $b$ quarks,
such as the exotic isosinglet quarks $h$, with improved sensitivity
in Run II of the Tevatron and with greatly expanded reach at the LHC.

\section*{Acknowledgments}
I wish to thank my collaborators on some of the topics mentioned here:
Troy Andre, Cheng-Wei Chiang, Michael Gronau, Zumin Luo, and Denis Suprun.
Michael Gronau and Hassan Jawahery also made helpful comments on the manuscript.
This work was supported in part by the United States Department of Energy
under Grant No.\ DE FG02 90ER40560.

\def \ajp#1#2#3{Am.\ J. Phys.\ {\bf#1}, #2 (#3)}
\def \apny#1#2#3{Ann.\ Phys.\ (N.Y.) {\bf#1}, #2 (#3)}
\def \app#1#2#3{Acta Phys.\ Polonica {\bf#1}, #2 (#3)}
\def \arnps#1#2#3{Ann.\ Rev.\ Nucl.\ Part.\ Sci.\ {\bf#1}, #2 (#3)}
\def \art{and references therein}
\def \cmts#1#2#3{Comments on Nucl.\ Part.\ Phys.\ {\bf#1} (#3) #2}
\def \cn{Collaboration}
\def \cp89{{\it CP Violation,} edited by C. Jarlskog (World Scientific,
Singapore, 1989)}
\def \econf#1#2#3{Electronic Conference Proceedings {\bf#1}, #2 (#3)}
\def \efi{Enrico Fermi Institute Report No.\ }
\def \epjc#1#2#3{Eur.\ Phys.\ J.\ C {\bf#1} (#3) #2}
\def \f79{{\it Proceedings of the 1979 International Symposium on Lepton and
Photon Interactions at High Energies,} Fermilab, August 23-29, 1979, ed. by
T. B. W. Kirk and H. D. I. Abarbanel (Fermi National Accelerator Laboratory,
Batavia, IL, 1979}
\def \hb87{{\it Proceeding of the 1987 International Symposium on Lepton and
Photon Interactions at High Energies,} Hamburg, 1987, ed. by W. Bartel
and R. R\"uckl (Nucl.\ Phys.\ B, Proc.\ Suppl., vol. 3) (North-Holland,
Amsterdam, 1988)}
\def \ib{{\it ibid.}~}
\def \ibj#1#2#3{~{\bf#1} (#3) #2}
\def \ichep72{{\it Proceedings of the XVI International Conference on High
Energy Physics}, Chicago and Batavia, Illinois, Sept. 6 -- 13, 1972,
edited by J. D. Jackson, A. Roberts, and R. Donaldson (Fermilab, Batavia,
IL, 1972)}
\def \ijmpa#1#2#3{Int.\ J.\ Mod.\ Phys.\ A {\bf#1} (#3) #2}
\def \ite{{\it et al.}}
\def \jhep#1#2#3{JHEP {\bf#1} (#3) #2}
\def \jpb#1#2#3{J.\ Phys.\ B {\bf#1}, #2 (#3)}
\def \lg{{\it Proceedings of the XIXth International Symposium on
Lepton and Photon Interactions,} Stanford, California, August 9--14, 1999,
edited by J. Jaros and M. Peskin (World Scientific, Singapore, 2000)}
\def \lkl87{{\it Selected Topics in Electroweak Interactions} (Proceedings of
the Second Lake Louise Institute on New Frontiers in Particle Physics, 15 --
21 February, 1987), edited by J. M. Cameron \ite~(World Scientific, Singapore,
1987)}
\def \kaon{{\it Kaon Physics}, edited by J. L. Rosner and B. Winstein,
University of Chicago Press, 2001}
\def \kdvs#1#2#3{{Kong.\ Danske Vid.\ Selsk., Matt-fys.\ Medd.} {\bf #1}, No.\
#2 (#3)}
\def \ky{{\it Proceedings of the International Symposium on Lepton and
Photon Interactions at High Energy,} Kyoto, Aug.~19-24, 1985, edited by M.
Konuma and K. Takahashi (Kyoto Univ., Kyoto, 1985)}
\def \mpla#1#2#3{Mod.\ Phys.\ Lett.\ A {\bf#1} (#3) #2}
\def \nat#1#2#3{Nature {\bf#1}, #2 (#3)}
\def \nc#1#2#3{Nuovo Cim.\ {\bf#1} (#3) #2}
\def \nima#1#2#3{Nucl.\ Instr.\ Meth.\ A {\bf#1}, #2 (#3)}
\def \np#1#2#3{Nucl.\ Phys.\ {\bf#1} (#3) #2}
\def \npps#1#2#3{Nucl.\ Phys.\ Proc.\ Suppl.\ {\bf#1} (#3) #2}
\def \npbps#1#2#3{Nucl.\ Phys.\ B Proc.\ Suppl.\ {\bf#1} (#3) #2}
\def \os{XXX International Conference on High Energy Physics, Osaka, Japan,
July 27 -- August 2, 2000}
\def \PDG{Particle Data Group, K. Hagiwara \ite, \prd{66}{010001}{2002}}
\def \pisma#1#2#3#4{Pis'ma Zh.\ Eksp.\ Teor.\ Fiz.\ {\bf#1}, #2 (#3) [JETP
Lett.\ {\bf#1}, #4 (#3)]}
\def \pl#1#2#3{Phys.\ Lett.\ {\bf#1} (#3) #2}
\def \pla#1#2#3{Phys.\ Lett.\ A {\bf#1}, #2 (#3)}
\def \plb#1#2#3{Phys.\ Lett.\ B {\bf#1} (#3) #2}
\def \pr#1#2#3{Phys.\ Rev.\ {\bf#1} (#3) #2}
\def \prc#1#2#3{Phys.\ Rev.\ C {\bf#1} (#3) #2}
\def \prd#1#2#3{Phys.\ Rev.\ D {\bf#1} (#3) #2}
\def \prl#1#2#3{Phys.\ Rev.\ Lett.\ {\bf#1} (#3) #2}
\def \prp#1#2#3{Phys.\ Rep.\ {\bf#1} (#3) #2}
\def \ptp#1#2#3{Prog.\ Theor.\ Phys.\ {\bf#1} (#3) #2}
\def \rmp#1#2#3{Rev.\ Mod.\ Phys.\ {\bf#1} (#3) #2}
\def \rp#1{~~~~~\ldots\ldots{\rm rp~}{#1}~~~~~}
\def \si90{25th International Conference on High Energy Physics, Singapore,
Aug. 2-8, 1990}
\def \slc87{{\it Proceedings of the Salt Lake City Meeting} (Division of
Particles and Fields, American Physical Society, Salt Lake City, Utah, 1987),
ed. by C. DeTar and J. S. Ball (World Scientific, Singapore, 1987)}
\def \slac89{{\it Proceedings of the XIVth International Symposium on
Lepton and Photon Interactions,} Stanford, California, 1989, edited by M.
Riordan (World Scientific, Singapore, 1990)}
\def \smass82{{\it Proceedings of the 1982 DPF Summer Study on Elementary
Particle Physics and Future Facilities}, Snowmass, Colorado, edited by R.
Donaldson, R. Gustafson, and F. Paige (World Scientific, Singapore, 1982)}
\def \smass90{{\it Research Directions for the Decade} (Proceedings of the
1990 Summer Study on High Energy Physics, June 25--July 13, Snowmass, Colorado),
edited by E. L. Berger (World Scientific, Singapore, 1992)}
\def \tasi{{\it Testing the Standard Model} (Proceedings of the 1990
Theoretical Advanced Study Institute in Elementary Particle Physics, Boulder,
Colorado, 3--27 June, 1990), edited by M. Cveti\v{c} and P. Langacker
(World Scientific, Singapore, 1991)}
\def \yaf#1#2#3#4{Yad.\ Fiz.\ {\bf#1}, #2 (#3) [Sov.\ J.\ Nucl.\ Phys.\
{\bf #1}, #4 (#3)]}
\def \zhetf#1#2#3#4#5#6{Zh.\ Eksp.\ Teor.\ Fiz.\ {\bf #1}, #2 (#3) [Sov.\
Phys.\ - JETP {\bf #4}, #5 (#6)]}
\def \zpc#1#2#3{Zeit.\ Phys.\ C {\bf#1}, #2 (#3)}
\def \zpd#1#2#3{Zeit.\ Phys.\ D {\bf#1}, #2 (#3)}

\end{document}